\documentclass[12pt]{article}
\usepackage{geometry}
\usepackage{amsmath, amssymb, graphicx,fullpage,color,mathtools,amsthm,xcolor}
\usepackage{caption, subcaption}
\usepackage{algorithm, algorithmic}
\usepackage{cite}
\usepackage[normalem]{ulem}

\newcommand{\mbf}{\mathbf}

\newcommand{\ang}[3]{\angle(#1,#2,#3)}
\newcommand{\bmat}[4]{\begin{bmatrix}#1 & #2 \\ #3 & #4\end{bmatrix}}

\usepackage{microtype}
\usepackage{hyperref,color}

\definecolor{webgreen}{rgb}{0,.35,0}
\definecolor{webbrown}{rgb}{.6,0,0}
\definecolor{RoyalBlue}{rgb}{0,0,0.9}
\definecolor{purp}{rgb}{0.6,0.05,0.8}
\definecolor{ora}{rgb}{0.7,0.35,0.02}

\hypersetup{
   colorlinks=true, linktocpage=true, 
   urlcolor=webbrown, linkcolor=RoyalBlue, citecolor=webgreen,
   pdfauthor={Levi H. Dudte, Gary P. T. Choi, Kaitlyn P. Becker, L. Mahadevan},
   pdfsubject={An additive framework for kirigami design}
}

\begin{document}

\author{Levi H. Dudte$^{1,\dagger}$, Gary P. T. Choi$^{2,\dagger}$, Kaitlyn P. Becker$^{1,3}$, L. Mahadevan$^{1,4,\ast}$\\
\\
\footnotesize{$^{1}$School of Engineering and Applied Sciences, Harvard University, Cambridge, MA, USA}\\
\footnotesize{$^{2}$Department of Mathematics, Massachusetts Institute of Technology, Cambridge, MA, USA}\\
\footnotesize{$^{3}$Department of Mechanical Engineering, Massachusetts Institute of Technology, Cambridge, MA, USA}\\
\footnotesize{$^{4}$Departments of Physics, and Organismic and Evolutionary Biology,}\\ \footnotesize{Harvard University, Cambridge, MA, USA}\\
\footnotesize{$^\dagger$L.H.D. and G.P.T.C. contributed equally to this work.}\\
\footnotesize{$^\ast$To whom correspondence should be addressed; E-mail: lmahadev@g.harvard.edu}
}
\title{An additive framework for kirigami design}
\date{}
\maketitle

\begin{abstract}
We present an additive approach for the inverse design of kirigami-based mechanical metamaterials by focusing on the empty (negative) spaces instead of the solid tiles. By considering each negative space as a four-bar linkage, we identify a simple recursive relationship between adjacent linkages, yielding an efficient method for creating kirigami patterns. This allows us to solve the kirigami design problem using elementary linear algebra, with compatibility, reconfigurability and rigid-deployability encoded into an iterative procedure involving simple matrix multiplications. The resulting linear design strategy circumvents the solution of a non-convex global optimization problem and allows us to control the degrees of freedom in the deployment angle field, linkage offsets and boundary conditions. We demonstrate this by creating a large variety of rigid-deployable, compact, reconfigurable kirigami patterns. We then realize our kirigami designs physically using two simple but effective fabrication strategies with very different materials. Altogether, our additive approaches present routes for efficient mechanical metamaterial design and fabrication based on ori/kirigami art forms.
\end{abstract}

\section*{Introduction}
Kirigami and origami are ancient art forms that have been rejuvenated as paradigms of a class of structural mechanical metamaterials derived from thin sheets. Kirigami works with cuts to create a globally coordinated, articulated, deployable structure that can lead to spectacular patterns in two and three dimensions, while origami works with creases to create a similar effect, although almost always in three dimensions. Both kirigami and origami can lead to complex structures that have been studied extensively in art, mathematics, science and engineering~\cite{callens2018flat,zhai2021mechanical}. However, the research literature on these subjects has had only a few points of convergence. Indeed, both artistic and technical books on origami~\cite{lang2012origami,demaine2007geometric,hull2020origametry} scarcely, if ever, mention, kirigami, and vice versa~\cite{temko1962kirigami}. From a mathematical perspective, although there are deep similarities between kirigami and origami, there are some subtle differences that require qualitatively different approaches to understand the implications of the local constraints on the global properties that result. In particular, both origami and kirigami exploit coordinated rotations of individual facets to achieve collective geometric changes of the connected structures; that is, they use piecewise deformations (with strong discontinuities along cuts/creases) to affect complex shape-shifting abilities. However, the nature of the discontinuity differs between origami and kirigami. In origami, creases are locations identified with a discontinuous geometry (slope jump), whereas in kirigami, cuts are identified with a topological discontinuity (position jump). \\

Numerous studies have explored different kirigami-based deployable structures, with a focus that is primarily on the forward problem, that of predicting the geometric and mechanical response of given designs~\cite{borcea2015geometric,chen2016topological,bertoldi2017flexible,celli2018shape,singh2021design,zheng2021continuum,czajkowski2022conformal}. The solution of the inverse problem typically involves setting up a global constrained optimization problem that encodes the geometric constraints in terms of the angles and lengths of the constituent tiles~\cite{konakovic2018rapid,choi2019programming,jiang2020freeform,choi2021compact,dang2021theorem}. Although such an approach is capable of producing patterns and mechanisms with different shape changes, the process of finding a solution to the nonlinear global constrained optimization problem for large patterns is both computationally difficult and hard to control. Inspired by biological morphogenesis driven by growing matter~\cite{thompson1942growth}, and its synthetic analog, additive manufacturing~\cite{truby2016printing}, one is naturally led to the question of whether it is possible to simplify the design of origami/kirigami metamaterials by switching from a global optimization perspective to some locally constrained additive process. In recent work, we showed that the use of local marching rules allows for an additive origami design approach~\cite{dudte2021additive}, converting a difficult non-convex global optimization problem~\cite{dudte2016programming} into a simpler, but still nonlinear, local problem. \\

In this Article we show that kirigami also lends itself to a simple additive approach that identifies and exploits a \emph{linear} marching construction to connect local growth rules to a global form. The subtle differences in the nature of the geometric and topological constraints between origami and kirigami imply that there are qualitative differences in these marching algorithms that lead to a much simpler algorithm for kirigami, despite it being a far richer paradigm for structural material design. Furthermore, we show that we can control the flexibility of encoding contractibility, compact reconfigurability and rigid-deployability for the design of quad kirigami in any contracted or deployed space using a simple design matrix characterized by a combination of certain edge-length and angle parameters. This substantially simplifies the process of designing kirigami patterns as matrix multiplications. Our approach of decoupling the length field, angle field and the boundary node constraints allows us to systematically analyze the degrees of freedom in the design space of different types of quad kirigami tessellations. \\

Finally, to realize these designs in practice, we describe two manufacturing approaches using additive manufacturing and casting/molding for robust kirigami structures with soft, fabric hinges.

\section*{Results}
\subsection*{Additive design framework}
We tackle the kirigami design problem by working our way up a ladder of hierarchical complexity, beginning with the smallest building block, a single four-bar linkage representing the negative space of a unit cell of four quads, before moving on to strips of linkages and finally linkage arrays. At each level, our strategy will be to describe a geometric marching construction, the \emph{forward} process, which can be cast as matrix multiplication, allowing us then to convert the marching algorithm into a flexible and intuitive design framework using linear \emph{inverse} design techniques.\\

In Fig.~\ref{fig:F1} we present an overview of the marching construction for quad kirigami design. The key idea is to identify the constraints in the negative spaces in a deployed configuration, and use them to form a design matrix that is nonlinearly dependent on the deployment angles and edge lengths. However, once the matrix is determined, the coordinates of all nodes can be obtained by a simple matrix multiplication step, yielding a linear design method for quad kirigami.

\subsubsection*{Linkage design}
To simplify our discussion, we start by considering the design of a negative space surrounded by four quads represented as a parallelogram four-bar linkage in $\mathbb{R}^2$ given by four points $\mbf{x}_k = (x_k, y_k)$, where $k = 0, 1, 2, 3$, with the \emph{deployment angle} $\phi = \ang{\mbf{x}_1}{\mbf{x}_0}{\mbf{x}_3}$ at $\mbf{x}_0$ (Fig.~\ref{fig:F1}A). We parameterize its construction according to 
\begin{equation}
\begin{bmatrix} I - Q & Q \\ -Q & I + Q \end{bmatrix}\begin{bmatrix} \mbf{x}_0 \\ \mbf{x}_3 \end{bmatrix} = \begin{bmatrix} \mbf{x}_1 \\ \mbf{x}_2\end{bmatrix},
\label{eq:link}
\end{equation}
where $Q = (1+\epsilon)R(-\phi)$ is a scaled rotation matrix, with $\epsilon$ being a scalar which we refer to as the \emph{offset}, $I = \bmat{1}{0}{0}{1}$ and $R(-\phi) = \bmat{\cos\-\phi}{\sin\-\phi}{-\sin\-\phi}{\cos\-\phi}$ are the identity and rotation matrices in $\mathbb{R}^2$, respectively, and $\begin{bmatrix}\mbf{x}_i\\\mbf{x}_j\end{bmatrix} = \begin{bmatrix}x_i\\y_i\\x_j\\y_j\end{bmatrix}$ is a column vector of coordinates of two points in the parallelogram. Extending Eq.~\eqref{eq:link} yields an $8\times4$ linkage design matrix $D$ that takes two points $\mbf{x}_0$ and $\mbf{x}_3$ (the red nodes in Fig.~\ref{fig:F1}A) as input and returns as output four points $\mbf{x}_k = (x_k, y_k)$, $k\in\{0,1,2,3\}$ to form the parallelogram:
\begin{equation}
D
\begin{bmatrix}
\mbf{x}_0 \\ \mbf{x}_3
\end{bmatrix} = 
\begin{bmatrix}
I & 0 \\ I - Q & Q \\ -Q & I + Q \\ 0 & I
\end{bmatrix}
\begin{bmatrix}
\mbf{x}_0 \\ \mbf{x}_3
\end{bmatrix} = 
\begin{bmatrix}
\mbf{x}_0 \\ \mbf{x}_1 \\ \mbf{x}_2 \\ \mbf{x}_3
\end{bmatrix}.
\label{eq:linkage}
\end{equation}
Our choice of parameterization measures a kind of eccentricity of the parallelogram (its departure from a rhombus) by the offset $\epsilon$ and forms a counter-clockwise polar coordinate system $(r = 1 + \epsilon, -\phi)$ centered at $\mbf{x}_0$ with $\phi=0$ in the direction of $\mbf{x}_3 - \mbf{x}_0$. When $\epsilon = 0$ the parallelogram is a rhombus and when $\epsilon=-1$ the parallelogram degenerates to two equal collinear line segments with $\mbf{x}_0=\mbf{x}_1$ and $\mbf{x}_2=\mbf{x}_3$. Otherwise, so long as $\phi\ne0,\pi$ the generated points $\mbf{x}$ form a parallelogram in $\mathbb{R}^2$ for all values of $\epsilon$. The points are ordered counter-clockwise when $\phi<\pi$, clockwise when $\phi>\pi$ and are collinear when $\phi=\pi$. Holding $\epsilon$ constant keeps each edge length in the parallelogram constant, while varying $\phi$ casts the parallelogram as a four-bar linkage. This allows for a one-dimensional deployment path, whereby two of the opposite interior angles in the linkage are $\phi$ and the other two are $\pi - \phi$ (see Supplementary Section 1.1 for more details). There are two major advantages for using parallelogram negative spaces. First, as described above, parallelogram slits have a simple parameterization involving only one parameter $\phi$ and one offset parameter $\epsilon$, substantially simplifying the final formulation for the kirigami design problem for large linkage arrays. Second, using parallelogram slits makes it easy to create rigid-deployable patterns that can morph from a closed and compact state to any deployed state, as we will discuss in the Methods section.

\subsubsection*{Linkage strips and linkage arrays}
Now we will analyze the design of two parallelogram four-bar linkages having one common point and generalize this system to strips and arrays of connected linkages. Consider the first two adjacent linkages in the first row of Fig.~\ref{fig:F1}B comprised of seven total nodes so that one node is shared by both. The linkage points can be labeled $\mbf{x}_{j,k}^0$ where $j\in\{0,1\}$ is the linkage index, $k\in\{0,1,2,3\}$ is the point in linkage index and $\mbf{x}_{0,2}^0=\mbf{x}_{1,0}^0$ denotes the shared node. The linkages each have their own deployment angles $\phi_j^0$ and offsets $\epsilon_j^0$. It can be observed that once $\mbf{x}_{0,0}^0$, $\mbf{x}_{0,3}^0$, $\phi_0^0$ and $\epsilon_0^0$ in the first linkage are given, $\mbf{x}_{0,1}^0$, $\mbf{x}_{0,2}^0$ are uniquely determined. In particular, as $\mbf{x}_{0,2}^0$ is already determined, it suffices to prescribe $\mbf{x}_{1,3}^0$, $\phi_1^0$ and $\epsilon_1^0$ to uniquely determine all points in the second linkage. \\

We can apply this insight to a more complex linkage strip of $n$ linkages $\mbf{x}_{j,k}^0$ as shown in the first row of Fig.~\ref{fig:F1}B, where $j\in\{0,\dots,n-1\}$ is the linkage index, $k\in\{0,1,2,3\}$ is a point in linkage index and $\mbf{x}_{j,2}^0=\mbf{x}_{j+1,0}^0$ denotes a shared node. In other words, we can obtain a series of linkage design matrices $D_0^0, D_1^0, \dots, D_{n-1}^0$ that encode the recursive dependency of the nodes on the previously determined nodes via dynamic programming (see Supplementary Section 1.2). Once the relationship in a linkage strip is determined, we can proceed to consider the relationship in a linkage array consisting of multiple linkage strips analogously, noticing that the $(i,j)$-th linkage in the array is dependent of all linkages $(\tilde{i},\tilde{j})$ with $\tilde{i} \leq i$ and $\tilde{j} \leq j$ recursively (see Supplementary Section 1.3 for more details). More specifically, the entire linkage array is dependent of the seed nodes $\{\mbf{x}_{0,0}^0, \mbf{x}_{0,3}^0, \mbf{x}_{1,3}^0, \dots, \mbf{x}_{n-1,3}^0, \mbf{x}_{0,0}^{1}, \mbf{x}_{0,0}^{2}, \dots, \mbf{x}_{0,0}^{m-1}\}$ consisting of all the left boundary points and all the top boundary points in the linkage array (the red nodes in Fig.~\ref{fig:F1}B), the deployment angle field $\{\phi_j^i\}$, and the offset field $\{\epsilon_j^i\}$. We remark that the deployment angles $\{\phi_j^i\}$ have to satisfy certain additional conditions in order to yield a kirigami pattern with different physical properties such as contractibility and compact reconfigurability, which will be discussed in the Methods section.

\subsubsection*{Global linear inverse design via matrix operations}
Putting all of these pieces together, for an array of $m \times n$ planar parallelogram four-bar linkages with nodes $\{\mbf{x}_{j,k}^{i}\}, i \in \{0,1,\dots,m-1\}, j \in \{0,1,\dots,n-1\}, k \in \{0,1,2,3\}$, we obtain a full linkage array design matrix $M$ of size $(4mn+2m+2n) \times (2m+2n)$. In particular, we have
\begin{equation}\label{eq:design_array}
 M\begin{bmatrix}\mbf{x}_{0,0}^0 \\ \mbf{x}_{0,3}^0 \\ \mbf{x}_{1,3}^0 \\ \vdots \\ \mbf{x}_{n-1,3}^0 \\ \mbf{x}_{0,0}^{1} \\ \mbf{x}_{0,0}^{2}  \\ \vdots  \\ \mbf{x}_{0,0}^{m-1}\end{bmatrix} =
\begin{bmatrix}
D_{0}^0 \\ 
D_{1}^0 \\
\vdots\\
D_{n-1}^0 \\ 
D_{0}^1 \\ 
D_{1}^1 \\ 
\vdots\\
D_{n-2}^{m-1} \\ 
D_{n-1}^{m-1}
\end{bmatrix} 
\begin{bmatrix}\mbf{x}_{0,0}^0 \\ \mbf{x}_{0,3}^0 \\ \mbf{x}_{1,3}^0 \\ \vdots \\ \mbf{x}_{n-1,3}^0 \\ \mbf{x}_{0,0}^{1} \\ \mbf{x}_{0,0}^{2}  \\ \vdots  \\ \mbf{x}_{0,0}^{m-1}\end{bmatrix} \\
 = \begin{bmatrix}\mbf{x}_{0,0}^0 \\ \mbf{x}_{0,1}^0 \\ \mbf{x}_{0,2}^0 \\ \mbf{x}_{0,3}^0 \\ \mbf{x}_{1,1}^0 \\ \mbf{x}_{1,2}^0 \\ \mbf{x}_{1,3}^0 \\ \vdots \\ \mbf{x}_{n-1,1}^{m-1} \\ \mbf{x}_{n-1,2}^{m-1} \\ \mbf{x}_{n-1,3}^{m-1} \end{bmatrix}
\end{equation}
with
\begin{equation}
M = \begin{bmatrix}
D_{0}^0 \\ 
D_{1}^0 \\
\vdots\\
D_{n-1}^0 \\ 
D_{0}^1 \\ 
D_{1}^1 \\ 
\vdots\\
D_{n-2}^{m-1} \\ 
D_{n-1}^{m-1}
\end{bmatrix} =
\begin{bmatrix}
I & 0 & 0 & 0 \\ 
G^{0,0}_{0,0} & G^{0,1}_{0,0} & 0 & 0 \\
G^{1,0}_{0,0} & G^{1,1}_{0,0} & 0 & 0 \\
0 & I & 0 & 0 \\
G^{0,0}_{0,1} G^{0,0}_{0,0} & G^{0,0}_{0,1} G^{0,1}_{0,0}  & G^{0,1}_{0,1} & 0 & \cdots \\
G^{1,0}_{0,1} G^{1,0}_{0,0} & G^{1,0}_{0,1} G^{1,1}_{0,0}  & G^{1,1}_{0,1} & 0 \\
0 & 0  & I & 0 \\
G^{0,0}_{0,2} G^{1,0}_{0,1} G^{1,0}_{0,0} & G^{0,0}_{0,2} G^{1,0}_{0,1} G^{1,1}_{0,0} & G^{0,0}_{0,2} G^{1,1}_{0,1} & G^{0,1}_{0,2} \\
G^{1,0}_{0,2} G^{1,0}_{0,1} G^{1,0}_{0,0} & G^{1,0}_{0,2} G^{1,0}_{0,1} G^{1,1}_{0,0} & G^{1,0}_{0,2} G^{1,1}_{0,1} & G^{1,1}_{0,2} \\ 
0 & 0  & 0 & I \\
& \vdots & & & \ddots \\
\end{bmatrix},
\end{equation}
where $D_{j}^{i}$ are the linkage design matrices which can be expressed in terms of the following generator matrices using dynamic programming (see Supplementary Section 1.3 for more details):
\begin{equation}
\begin{bmatrix}
G^{0,0}_{i,j} & G^{0,1}_{i,j} \\ G^{1,0}_{i,j} & G^{1,1}_{i,j}
\end{bmatrix}=
\begin{bmatrix}
I - (1+\epsilon_j^i)R(-\phi_j^i) & (1+\epsilon_j^i)R(-\phi_j^i) \\ -(1+\epsilon_j^i)R(-\phi_j^i) & I + (1+\epsilon_j^i)R(-\phi_j^i)
\end{bmatrix}.
\end{equation}

\begin{figure}[t]
\centering
\includegraphics[width=0.8\textwidth]{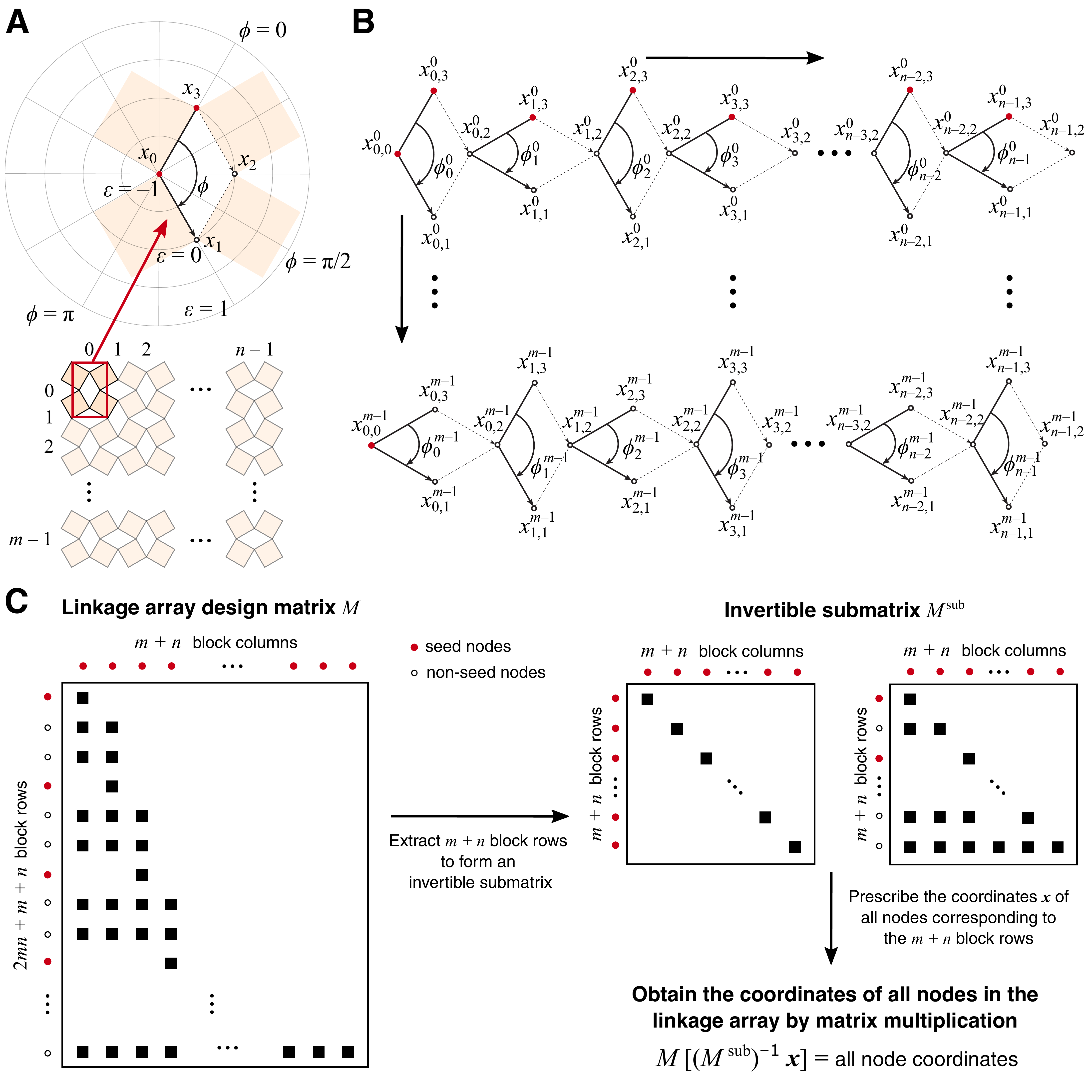}
\caption[]{\small{\textbf{Marching construction for quad kirigami design.} \textbf{(A)}~Geometric design of a four-bar linkage representing a negative space in an $(m+1)\times (n+1)$ quad kirigami pattern. Given the coordinates of two seed nodes $\mbf{x}_0$ and $\mbf{x}_3$ (red), the linkage can be parameterized by a deployment angle $\phi$ and an edge length parameter (offset) $\epsilon$. \textbf{(B)} Each linkage strip consists of a series of four-bar linkages each parameterized by $\phi_j$ and $\epsilon_j$. The coordinates of the non-seed nodes (hollow dots) are dependent of the seed nodes (red dots) as indicated by the arrows in the linkages. One can grow the linkages following the direction of the large arrows to obtain $m$ linkage strips each with $n$ linkages, thereby forming an $m \times n$ linkage array with nodes $\{\mbf{x}_{j,k}^{i}\}, i \in \{0,1,\dots,m-1\}, j \in \{0,1,\dots,n-1\}, k \in \{0,1,2,3\}$. \textbf{(C)} Considering the relationship between all nodes in the linkage array, we obtain a design matrix $M$ with all parameters $\{\phi_j^i\}$ and $\{\epsilon_j^i\}$ encoded in it (left). Note that $M$ consists of $2mn+m+n$ block rows and $m+n$ block columns, where each block (black square) is a $2\times 2$ matrix corresponding to the $xy$-coordinates of the each node (see also Eq.~\eqref{eq:design_array}). One can then extract $m+n$ block rows suitably to form a $(2m+2n)\times (2m+2n)$ invertible submatrix $M^{\text{sub}}$. In particular, choosing exactly the $m+n$ block rows corresponding to all seed nodes (red dots) will form an identity matrix $M^{\text{sub}} = I_{2m+2n}$ (middle). Alternatively, one can choose block rows corresponding to some seed nodes (red dots) and some non-seed nodes (hollow dots) to form $M^{\text{sub}}$ (right). In either case, we can then prescribe the coordinates of all nodes corresponding to the chosen $m+n$ block rows, and then invert the submatrix $M^{\text{sub}}$ to solve for the seed node coordinates. Finally, by a direct matrix multiplication of $M$ with the solved seed node coordinates, we obtain the entire linkage array.}}
\label{fig:F1}
\end{figure}

The sparsity of the matrix $M$ forms a simple pattern (Fig.~\ref{fig:F1}C): after the first block row, each of the next $n$ set of three block rows has entries in a new block column unoccupied by all block rows above it in the matrix. For example, the fifth, sixth and seventh block rows in Eq.~\eqref{eq:design_array} contain $G^{0,1}_{0,1}$, $G^{1,1}_{0,1}$ and $I$, respectively, as entries in the third block column, which contains only zero entries in the block rows above the fifth. After that, each of the next $(m-1)$ set of $(n+1)$ block rows has entries in a new block column unoccupied by all block rows above it in the matrix. This makes it easy to observe that a submatrix $M^\text{sub}$ of $M$ comprised of any two block rows from the first four block rows, and one block row from each of the subsequent sets of block rows (i.e. with a total of $2+(n-1)+(m-1) = m+n$ block rows) will be a $(2m+2n) \times (2m+2n)$ square matrix with full rank and hence invertible.\\

This observation suggests a linear design strategy for quad kirigami patterns. To create a quad kirigami pattern consisting of $(m+1) \times (n+1)$ quads, we need to fully determine an array of $m \times n$ planar parallelogram four-bar linkages representing all negative spaces. This linear inverse design process for the linkage array is summarized as follows:
\begin{enumerate}
\item Assemble the full linkage array design matrix $M$ (Fig.~\ref{fig:F1}C, first panel). Note that this requires choosing the deployment angle $\phi_j^i$ and offset $\epsilon_j^i$ for each linkage $(i,j)$, which will be discussed in detail in the next section.
\item Choose a subset of $(m+n)$ points $\{\mbf{\tilde{x}}_0, \mbf{\tilde{x}}_1, \dots, \mbf{\tilde{x}}_{m+n-1}\}$ in the linkage array such that the corresponding block rows of them in $M$ form a $(2m+2n)\times (2m+2n)$ matrix $M^\text{sub}$ with full rank (Fig.~\ref{fig:F1}C, second panel). The coordinates of these points can be prescribed directly by the designer as boundary conditions. We can then use them to find the seed nodes in Eq.~\eqref{eq:design_array}:
\begin{equation}
 \begin{bmatrix}\mbf{x}_{0,0}^0 \\ \mbf{x}_{0,3}^0 \\ \mbf{x}_{1,3}^0 \\ \vdots \\ \mbf{x}_{n-1,3}^0 \\ \mbf{x}_{0,0}^{1} \\ \mbf{x}_{0,0}^{2}  \\ \vdots  \\ \mbf{x}_{0,0}^{m-1}\end{bmatrix} =  (M^\text{sub})^{-1} \begin{bmatrix} \mbf{\tilde{x}}_0 \\ \mbf{\tilde{x}}_1 \\ \mbf{\tilde{x}}_2 \\\vdots \\ \mbf{\tilde{x}}_{m+n-2} \\ \mbf{\tilde{x}}_{m+n-1}\end{bmatrix}.
 \label{eq:solved_seed}
\end{equation}
In particular, if we simply choose $\{\mbf{\tilde{x}}_0, \mbf{\tilde{x}}_1, \dots, \mbf{\tilde{x}}_{m+n-1}\}$ to be the seed nodes (i.e. all the top and left boundary points in the linkage array), then $M^\text{sub}$ is simply the $(2m+2n) \times (2m+2n)$ identity matrix.
\item Calculate the full linkage array by a direct matrix multiplication using Eq.~\eqref{eq:solved_seed} together with Eq.~\eqref{eq:design_array}  (Fig.~\ref{fig:F1}C, third panel).
\item For the boundary nodes of the $(m+1) \times (n+1)$ kirigami pattern that are not included in the linkage array, one can determine their coordinates uniquely using the linkage array obtained in the last step together with four prescribed corner positions and a set of desired \emph{boundary offsets} (see Supplementary Section 1.4 for details).
\end{enumerate}
\subsubsection*{Choices of the deployment angles and the offsets}
As discussed in the above subsection on linkage strip and linkage arrays, the full design matrix $M$ is determined by the deployment angles and offsets of every four-bar linkage negative space. Here, the angles $\{\phi_j^i\}$ in the negative spaces are related by certain local angle rules. Depending on the desired properties of the resulting kirigami pattern, the degrees of freedom (DOFs) in them will be different. In particular, note that the angles of the tiles can be expressed in terms of the deployment angles (Fig.~\ref{fig:F2}A). Therefore, we can achieve contractibility~\cite{choi2019programming}, reconfigurability~\cite{choi2021compact} and rigid-deployability~\cite{choi2021compact} by enforcing additional constraints on the deployment angles $\{\phi_j^i\}$ (Fig.~\ref{fig:F2}B). The offsets $\{\epsilon_j^i\}$ of all linkages can be chosen independently as long as they do not lead to degeneracies or self-intersections (Fig.~\ref{fig:F2}C). More details are available in the Methods section.

\subsection*{Linear and nonlinear inverse design}
With the proposed additive design approach, we now have full control of the design of quad kirigami patterns. In particular, we have characterized the design space of contractible quad kirigami patterns as well as that of rigid-deployable, compact reconfigurable quad kirigami patterns: We can inverse design all such patterns by prescribing the desired interior properties via the deployment angle and offset fields in the design matrix, specifying certain boundary conditions, solving a matrix equation to find the corresponding seed coordinates, and finally using a direct matrix multiplication to obtain the resulting patterns. Moreover, as the deployment angle field and the boundary conditions are set in two separate steps, this linear design approach suggests that it is possible to control the target shape of a kirigami pattern (related to the boundary conditions) and the state of deployment (related to the deployment angle) separately. To illustrate this idea, in Fig.~\ref{fig:F3}A we design two rigid-deployable, compact reconfigurable kirigami patterns that deploy to a circle at the deployed state with $\phi = \pi/2$ (i.e. all negative spaces are rectangular) and a trapezium at the deployed state with $\phi = \pi/4$ (i.e. all negative spaces are parallelogram with the acute angle being $\pi/4$) respectively, from which it can be observed that we have precise control of the shape matching and the deployment angle (see Supplementary Section 3 for more results).\\

A natural next question is whether it is possible to match multiple target shapes at multiple stages of deployment. Since the boundary conditions have already been used for enforcing the shape of the pattern at a certain stage, we cannot use them to control the shape at another state. Nevertheless, there are still DOFs in the angles and edge lengths in setting the design matrix. This suggests a nonlinear inverse kirigami design approach for approximating multiple prescribed shapes at multiple states: Given a certain nonlinear objective function that quantifies the optimality of a kirigami pattern, we can search for an optimal pattern effectively by solving the optimization problem over the remaining set of the design parameters. For instance, we can solve for a target second contracted shape by fixing the deployment angle as $\phi = \pi$ and optimizing the offset parameters $\{\epsilon_j^i\}$ (see Supplementary Section 2 for more details). Fig.~\ref{fig:F3}B shows various nonlinear inverse design results of rigid-deployable, compact reconfigurable kirigami patterns that morph from a closed and compact square to a target second closed and compact shape. Analogous to our recent work~\cite{choi2021compact}, our method is capable of producing a square-to-circle shape change. Moreover, we can achieve a lot more different target shapes with different curvature properties at the second contracted state (see Supplementary Section 3 for more results). The larger variety of patterns we can obtain is attributed to the simplified parameter space in the proposed additive design formulation. \\

With the above linear and nonlinear inverse design results, one may ask about the limit of the shapes achievable by introducing cuts on a square. As the design space of the rigid-deployable, compact reconfigurable quad kirigami patterns is fully characterized by our framework, we can easily generate a large number of such patterns by setting the design matrix parameters randomly and perform a statistical analysis. Here we fix the linkage array size as $m\times n = 10\times 10$ and generate 10000 kirigami patterns on a unit square with random offsets $\{\epsilon_j^i\}$. More specifically, for each trial, the boundary nodes of the pattern are enforced to lie on a unit square with uniform spacing and the deployment angle $\phi$ is set to be 0. For the offset $\epsilon_j^i$ of each linkage $(i,j)$, we first randomly sample $r_j^i \in [-1,1]$ and then set $\epsilon_j^i = 10^{r_j^i}-1$, thereby ensuring that $\epsilon_j^i \in [-0.9, 9]$ and there is no degeneracy. We then use the proposed method with the above parameters to obtain a rigid-deployable, compact reconfigurable square kirigami pattern, and then assess the geometric properties of its fully deployed configuration ($\phi = \pi/2$) and its second contracted configuration ($\phi = \pi$) such as diagonal ratio and area as functions of $\phi$ (see Supplementary Section 4 for more details). Specifically, by considering the ratio $r_d$ between the two diagonal lengths of a structure at different states, we can assess the shear of the overall shape as $\phi$ increases. As shown in Fig.~\ref{fig:F3}C, the diagonal ratio at the fully deployed state ($r_d(\pi/2)$) and that at the second contracted state ($r_d(\pi)$) forms a highly linear relationship. It is also noteworthy that the reconfigured diagonal ratio is always larger than the deployed diagonal ratio, which suggests a natural limitation on the possible shape change achievable throughout the deployment. As for the overall area of the pattern $r_a$, it can be shown that the maximum deployed area of the pattern is always achieved at $\phi = \pi/2$. Comparing the diagonal ratio and the maximum deployed area of the random patterns, we find that these quantities are positively correlated. One can also consider other quantities such as the side length ratio $r_l$ and study the relationship between them (see Supplementary Section 4). We remark that as the proposed design framework only involves simple matrix operations, the generation and analysis of each pattern are highly efficient and take less than 0.5 seconds.

\subsection*{Physical model fabrication}
Realizing the rigid-deployable, compact reconfigurable kirigami patterns obtained by the additive design framework requires a careful treatment of the hinges in between the tiles to enable compact reconfigurability of the patterns in practice. While it is possible to use tape joints to connect the tiles as illustrated in our prior work~\cite{choi2021compact}, the fabrication is time-consuming and the hinges are likely to fatigue and break. To circumvent both issues, we propose two fabrication techniques for manufacturing robust kirigami models: an additive manufacturing (3D printing) method (Fig.~\ref{fig:F4}A--D) and a molded elastomeric composite method (Fig.~\ref{fig:F4}E--H). More details are available in the Methods section.

\section*{Discussion}
Our additive perspective for kirigami design circumvents the difficulty of non-convex optimization in favor of a much simpler recursive method based on linear algebra. By parameterizing each negative space by two parameters $\phi$ and $\epsilon$ and formulating the kirigami design problem as a matrix multiplication based on the relationship between adjacent linkages, our method decouples the fields of deployment angles and lengths of the cut patterns and hence gives a better understanding of the design space of quad kirigami. From the structure of the design matrix, one can also precisely identify the DOFs in the control of the overall pattern shape from the viewpoint of matrix rank. Altogether, our proposed formulation effectively characterizes the design space of quad kirigami patterns for a range of problems in science and engineering. Complementing this, our combination of 3D printing and casting approaches provides a relatively inexpensive and efficient way of manufacturing kirigami models with fatigue-resistant hinges. \\

A limitation of our work is that our approach currently only focuses on 2D quad kirigami design with parallelogram negative spaces. Possible future directions include extending the additive design formulation to more general kirigami patterns with non-parallelogram negative spaces~\cite{choi2019programming} or other polygons, as well as generalizations to non-planar linkages in 3D, which would require the consideration of additional geometric constraints.\\

\section*{Methods}

\subsection*{Choices of the deployment angles and the offsets}
\subsubsection*{Contractibility}
\begin{figure}[t]
\centering
\includegraphics[width=0.8\textwidth]{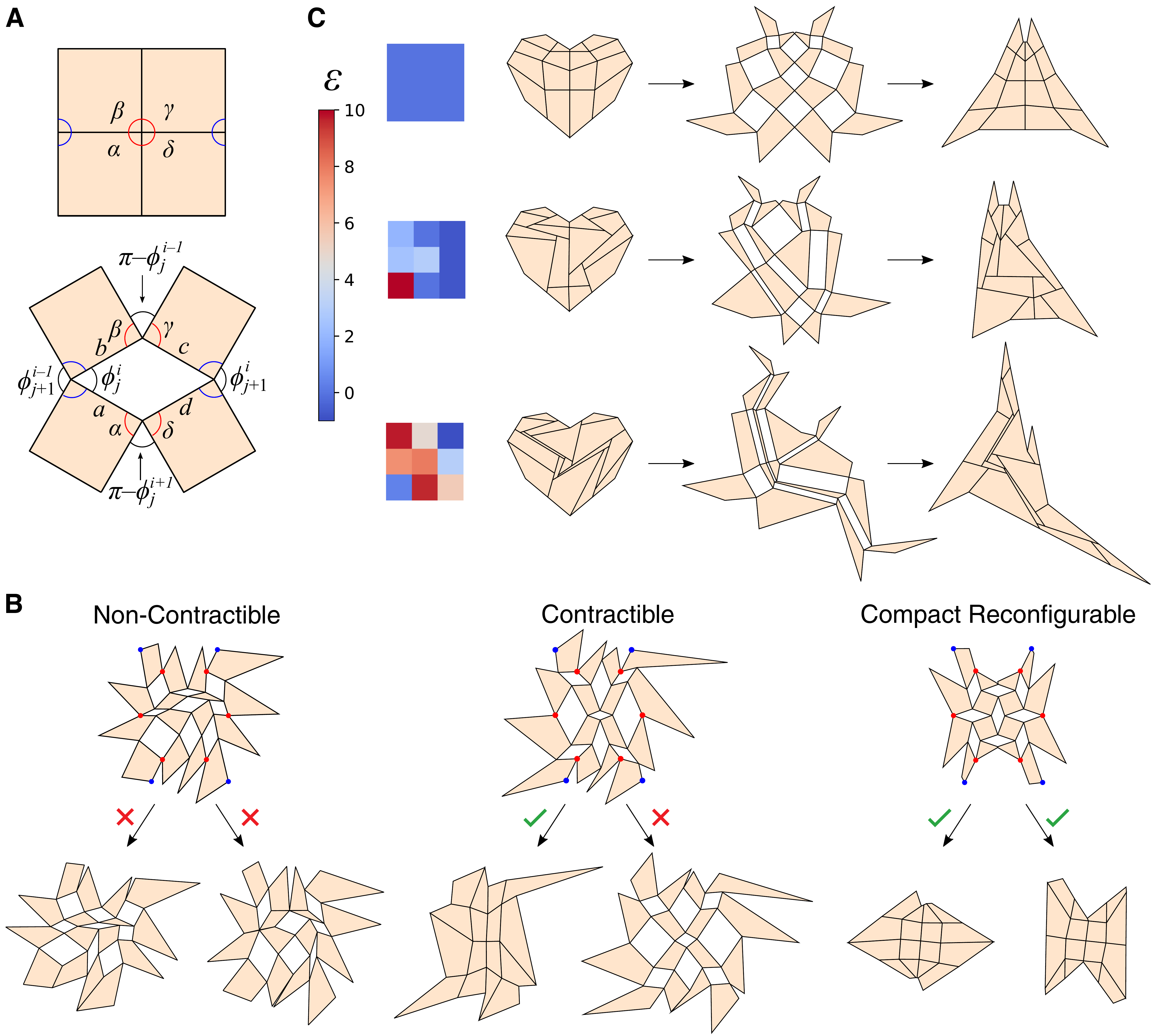}
\caption[]{\textbf{Encoding different desired properties in the design matrix.} \textbf{(A)} The angle constraints related to contractibility and compact reconfigurability can all be expressed in terms of the deployment angles of adjacent linkages and encoded in the design matrix. The top panel shows a contracted unit cell of four quads with the four angles $\alpha, \beta, \gamma, \delta$ at the center. The bottom panel shows a deployed configuration of the unit cell, where $\phi_j^i, \phi_{j-1}^i, \phi_{j+1}^i$ are the deployment angles of adjacent linkages. Note that the four angles $\alpha, \beta, \gamma, \delta$ highlighted in red determine the contractibility of the unit cell, while the four angles highlighted in blue determine the compact reconfigurability of it. \textbf{(B)} Example choices of the deployment angle field $\{\phi_j^i\}$ that lead to a non-contractible pattern (left), a contractible pattern (middle) and a compact reconfigurable pattern (right). The top row shows three patterns obtained using the proposed design method with a given deployment angle field, where the same linkage boundary node constraints (highlighted in red) and corner constraints (highlighted in blue) are used. In all examples, the offset field $\{\epsilon_j^i\}$ is set to be 0. The bottom row shows the two maximally contracted configurations of each pattern, where the red crosses indicate that the maximally contracted configurations are not closed and compact and the green check marks indicate that they are closed and compact. It can be observed that two closed and compact contracted configurations can be achieved only by using a deployment angle field satisfying the compact reconfigurability condition. \textbf{(C)}~Example choices of the offset field $\{\epsilon_j^i\}$ for the $3\times 3$ negative space in a $4\times 4$ kirigami pattern. Three compact reconfigurable, rigid-deployable heart shapes are obtained by solving the matrix equation, with the offset at each of $3\times 3$ linkages shown in the corresponding square indicators (see the blue/red gradient for the values of each $\epsilon$).}
\label{fig:F2}
\end{figure}

\begin{figure}[t]
\centering
\includegraphics[width=\textwidth]{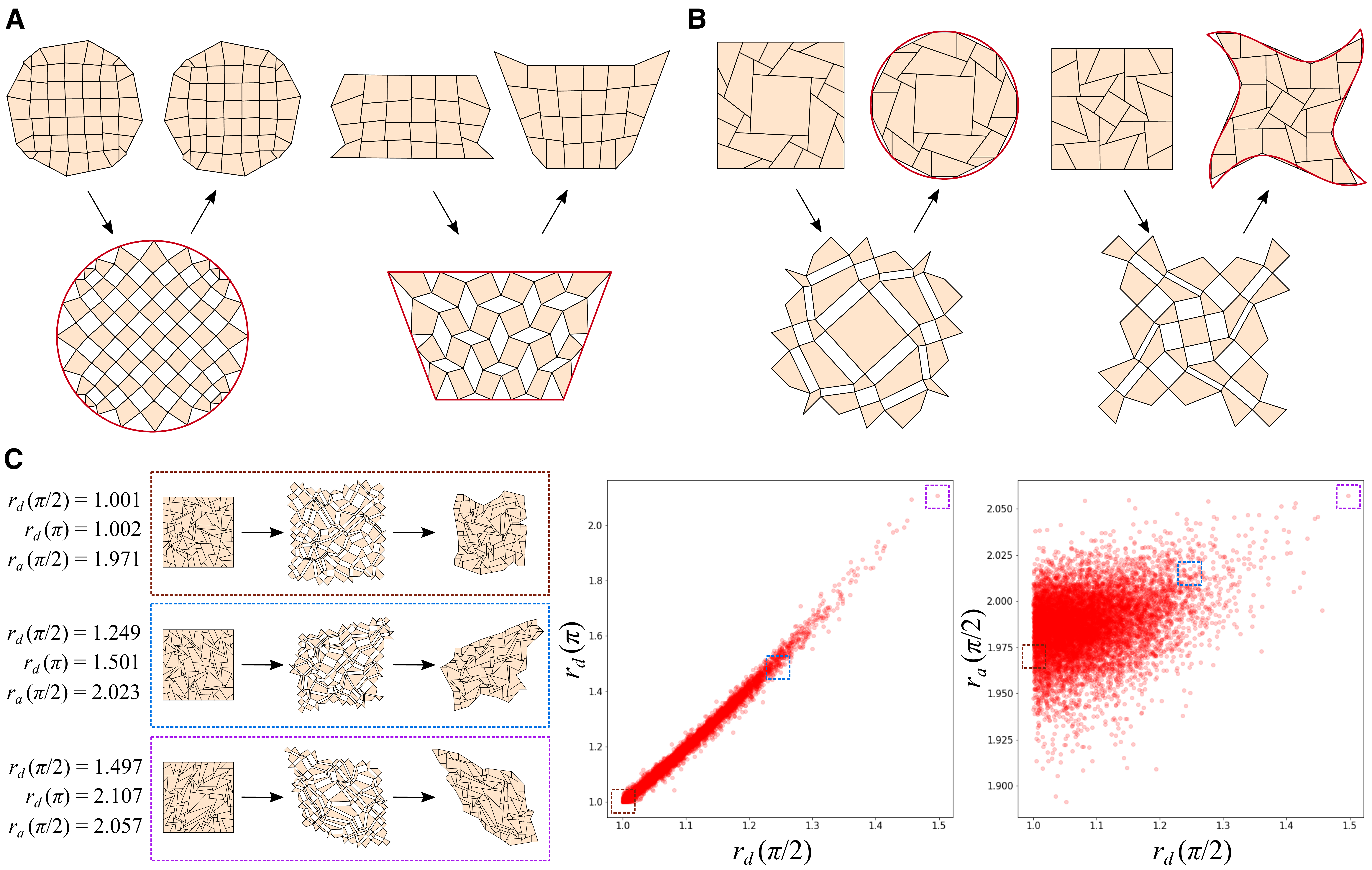}
\caption{\textbf{Linear and nonlinear inverse design of quad kirigami patterns using the proposed framework.} \textbf{(A)} Rigid-deployable, compact reconfigurable kirigami patterns that approximate a target circle and trapezium at two prescribed deployed states $\phi = \pi/2$ and $\phi = \pi/4$ respectively. \textbf{(B)}~Patterns that morph from a square to a circle (positive curvature) and a star shape (mixed curvature) in the second contracted state. \textbf{(C)} Statistical analysis of 10000 compact reconfigurable, rigid-deployable square kirigami patterns with random cuts. The left column shows the deployment paths of three random patterns. The middle and right plots show the relationship between the diagonal ratio at the fully deployed state $r_d(\pi/2)$, the diagonal ratio at the second contracted state $r_d(\pi)$ and the maximum deployed area $r_a(\pi/2)$ for all 10000 randomly generated patterns, where each red dot represents a pattern. The positions of the three example patterns on the two scatter plots are boxed.}
\label{fig:F3}
\end{figure}

For a quad kirigami pattern to be contractible, each unit cell consisting of four adjacent tiles has to satisfy an edge length constraint and an angle sum constraint~\cite{choi2019programming}. Specifically, suppose the four edges of a negative space are with length $a,b,c,d$ and the four angles of the tiles are $\alpha,\beta,\gamma,\delta$ as shown in Fig.~\ref{fig:F2}A, the following conditions must be satisfied:
\begin{equation}\label{eqt:edgelength}
    a+d = b+c,
\end{equation}
\begin{equation}\label{eqt:anglesum}
    \alpha+\beta+\gamma+\delta = 2\pi.
\end{equation}
In our formulation of the parallelogram four-bar linkages, we have $a = c$ and $b = d$ and hence the edge length constraint in Eq.~\eqref{eqt:edgelength} is automatically satisfied (note that here we do not require the four angles to meet at a common point as we have the flexibility to change the offset). Now, we rewrite the angle sum constraint using the deployment angles of the $(i,j)$-th four-bar linkage and its adjacent linkages. As shown in Fig.~\ref{fig:F2}A, we have
\begin{equation}
    \alpha+\delta + (\pi - \phi_j^i) + (\pi - \phi_j^{i+1}) = 2\pi,
\end{equation}
\begin{equation}
    \beta+\gamma + (\pi - \phi_j^i) + (\pi - \phi_j^{i-1}) = 2\pi.
\end{equation}
Therefore, the angle sum constraint in Eq.~\eqref{eqt:anglesum} can be rewritten as
\begin{equation}
    \phi_j^{i-1} + 2 \phi_j^i + \phi_j^{i+1} = 2\pi.
\end{equation}
As the deployment direction of adjacent linkages is alternating, for the entire $m \times n$ linkage array, we have
\begin{equation} \label{eqt:contractible_phi}
    \left\{\begin{array}{ll}
        \phi_j^{i-1} + 2 \phi_j^i + \phi_j^{i+1} = 2\pi & \text{ if } i+j \text{ is odd},\\
        \phi_{j-1}^i + 2 \phi_{j}^i + \phi_{j+1}^{i} = 2\pi & \text{ if } i+j \text{ is even},
    \end{array}\right.
\end{equation}
where $i = 0, \dots, m-1$ and $j =  0, \dots, n-1$. Note that some of the above equations involve ghost points with out-of-range indices, namely $\phi_j^{-1}, \phi_j^{m}, \phi_{-1}^{i}, \phi_{n}^{i}$, which are free variables that can be prescribed arbitrarily. Once the values of them have been prescribed, Eq.~\eqref{eqt:contractible_phi} has full rank and gives a unique set of $\phi_j^i$ values for all $i,j$. In other words, the number of DOFs in the space of deployment angles of a contractible $m\times n$ linkage array is exactly $2(m+n)$.  \\

Fig.~\ref{fig:F2}B (left) shows an example deployed pattern obtained by the linear inverse design approach, where the offset field $\{\epsilon_j^i\}$ is set to be uniformly 0 in the design matrix and the deployment angle field $\{\phi_j^i\}$ is set arbitrarily. The coordinates of certain linkage array boundary nodes (red) are prescribed as the boundary conditions for solving the seed coordinates to get the linkage array, and then the four corners of the kirigami pattern (blue) are further prescribed to uniquely determine the remaining boundary nodes of the resulting kirigami pattern. It can be observed that the resulting pattern is non-contractible, with the two maximally contracted states containing holes. By contrast, Fig.~\ref{fig:F2}B (middle) shows an example pattern obtained by the linear inverse design approach with the contractibility constraint in Eq.~\eqref{eqt:contractible_phi} enforced in the design matrix and all other parameters being the same as the previous example. This time, we can see that the resulting pattern admits a closed and compact contracted configuration.

\subsubsection*{Reconfigurability}
For a kirigami pattern to be compact reconfigurable, i.e. admitting two closed and compact contracted states, it should satisfy the above contractibility constraints as well as a set of dual edge length constraints $a+b = c+d$ and dual angle sum constraints~\cite{choi2021compact}. In the formulation of the parallelogram four-bar linkages, the dual edge length constraint is again automatically satisfied as we have $a = c$ and $b = d$ in Fig.~\ref{fig:F2}A. Note that the dual angle sum constraint at every four-bar linkage $(i,j)$ is simply the dual case in Eq.~\eqref{eqt:contractible_phi}. In other words, for a compact reconfigurable quad kirigami pattern, the corresponding linkage array should satisfy 
\begin{equation}\label{eqt:reconfigurable_phi}
    \left\{\begin{array}{ll}
        \phi_j^{i-1} + 2 \phi_j^i + \phi_j^{i+1} = 2\pi,\\
        \phi_{j-1}^i + 2 \phi_{j}^i + \phi_{j+1}^{i} = 2\pi,\\
    \end{array}\right.
\end{equation}
for every pair of $(i,j)\in \{0, 1, \dots, m-1\}\times \{0, 1, \dots, n-1\}$. This implies that
\begin{equation}\label{eqt:reconfigurable_phi2}
    \left\{\begin{array}{ll}
        \phi_j^i = \phi & \text{ if } i+j \text{ is even},\\
        \phi_j^i = \pi - \phi & \text{ if } i+j \text{ is odd},
    \end{array}\right.
\end{equation}
where $\phi$ is the deployment angle of the first linkage in the array. In other words, there is exactly 1 DOF in the entire deployment angle field $\{\phi_j^i\}$. In particular, this DOF captures the state of deployment of the pattern. Note that this agrees with the observation in our recent study~\cite{choi2021compact}. \\

Fig.~\ref{fig:F2}B (right) shows an example pattern obtained by solving the matrix equation with Eq.~\eqref{eqt:reconfigurable_phi} enforced, where the boundary condition is given by the same set of boundary linkage vertices (red) and corner vertices (blue) as in the previous examples. It can be observed that the resulting pattern admits multiple closed and compact contracted configurations. We also note that due to the difference in the flexibility of the angle field, the resulting patterns in Fig.~\ref{fig:F2}B are substantially different in shape. 

\subsubsection*{Rigid-deployability}
For the structure to be able to morph from a closed and compact contracted state to a deployed state with no tile strain, we need to satisfy certain conditions on the edge lengths and angles of the kirigami tiles~\cite{choi2021compact}. Adapting this idea to the current additive design framework, we note that rigid-deployability depends on the geometry of the parallelogram four-bar linkages as well as the contractibility and compact reconfigurability discussed above. For any non-contractible pattern, e.g. the case in Fig.~\ref{fig:F2}B (left), the pattern is always rigid-deployable in between the two maximally contracted states as the parallelogram four-bar linkages can deploy and contract without any geometrical frustration. For any contractible pattern, e.g. the case in Fig.~\ref{fig:F2}B (middle), rigid-deployability near the fully contracted state depends on the geometry of the four-bar linkages. If all four-bar linkages are not rhombi, they will always form a straight line at the fully contracted state and hence the kirigami pattern is rigid-deployable in between the fully contracted state and the second maximally contracted state. However, if some of the four-bar linkages are rhombi, they do not necessarily form a straight line at the fully contracted state and hence the pattern will generally be not rigid-deployable unless certain extra angle constraints are enforced as described in~\cite{choi2021compact}. In fact, it can be observed that the contractible kirigami pattern in Fig.~\ref{fig:F2}B (middle) contains negative spaces that do not form a straight line in the fully contracted state, which indicates that the pattern is not fully rigid-deployable but only rigid-deployable within a certain range of angles (see Supplementary Section 1.5 for a more detailed discussion).\\

For any compact reconfigurable pattern created by our approach such as Fig.~\ref{fig:F2}B (right), the parallelogram four-bar linkage assumption together with the angle constraint in Eq.~\eqref{eqt:reconfigurable_phi2} will automatically satisfy the conditions for rigid-deployability in~\cite{choi2021compact}. Therefore, the compact reconfigurable kirigami patterns created by our approach are always rigid-deployable.

\subsubsection*{Offset field}
We note that the offsets $\{\epsilon_j^i\}$ of all linkages can be chosen independently as long as they do not lead to degeneracies or self-intersections. As shown in Fig.~\ref{fig:F2}C, different choices of the offsets can be used for creating a compact reconfigurable, rigid-deployable heart structure. Setting a uniform offset $\epsilon_j^i = 0$ at all linkages leads to a more regular and symmetric second contracted state (top row), while setting a large offset $\epsilon_j^i \gg 1$ for one particular linkage creates a more irregular shape locally (middle row). Setting a large offset at multiple linkages produces a much more irregular second contracted state (bottom row). This naturally raises the question of what conditions would yield admissible kirigami patterns without self-intersection, just as in origami~\cite{lang2018rigidly}. A detailed analysis (see Supplementary Section 1.6) allows us to establish a theorem ensuring no self-intersections and also provide empirical insights relevant for practical applications.\\

\subsection*{Implementation}
The proposed framework is implemented in Python. Note that our framework involves solving a matrix equation to determine the deployment angle field $\{\phi_j^i\}$ (either using the contractible formulation or the compact reconfigurable formulation), and a matrix equation to find the seed positions in the linkage array from the prescribed node positions. These matrix equations are solved using the function \texttt{numpy.linalg.solve}. \\

For generating patterns that approximate a target shape, we compute the projection of the boundary points onto the target shape using the Python package \texttt{shapely} and evaluate the distance between them. The nonlinear optimization is done using Python's optimization and root finding toolbox \texttt{scipy.optimize}. We set all $\epsilon_j^i = 0$ as the initial condition and search for the optimal $\epsilon_j^i \in (-1, \infty)$ for all linkages. Empirically, we do not observe self-intersections in the optimization results even without imposing any additional constraints, possibly due to the fact that the occurrence of self-intersections is related to extreme values of $\epsilon_j^i$, which are far from the initial condition. However, if one wants to ensure that there is no self-intersection, it is possible to further restrict all $\epsilon_j^i$ to a smaller range or introduce additional inequality constraints enforcing that all tile and linkage orientations are consistent (by considering the direction of the outward normal vector obtained using the cross product of two edges).

\subsection*{Physical model fabrication}
We introduced two fabrication approaches for producing rigid-deployable kirigami structures, namely the 3D printing approach and the rubber molding approach. While fabrication is not limited to these two approaches, the purpose of introducing two methods is to give multiple viable options for construction that come with the trade-offs from variations in the fabrication strategy, as well as differing materials compatible with the two approaches. The 3D printing approach presented below can be expanded to incorporate more complex geometries more easily than the casting approach. The printed method also can be leveraged for more modular assembly strategies and inline customization. The molding approach can be modified to achieve a higher throughput than would be possible with the printing approach. Using a soft resin in the casting method can also enable some designs that must pass through geometric frustration and simplifies the mold design by not requiring draft angles and relief features. Casting could also be done with a more rigid resin, or a combination of rigid and soft materials.

\subsubsection*{Overview}
To illustrate the 3D printing method, we fabricate the heart pattern obtained by linear inverse design using this approach (Fig.~\ref{fig:F4}A--D). The rigid tiles were created by 3D printing with hollow slots where fatigue-resistant fabric hinges can be glued in place (Fig.~\ref{fig:F4}D). Although the tiles can be printed individually and arranged for maximal packing density on a printer bed,  we chose to print the tiles in a configuration halfway deployed between the two target design configurations in order to facilitate hinge insertion and balancing tension in the assembly. A vector file of the tile pattern is imported into a CAD software (Fusion360) and extruded to the desired thickness ($20~mm$ for the heart example in (Fig.~\ref{fig:F4}B). Slots of $0.75~mm$ in width are cut into the tiles of the CAD file to create space for the insertion of a fabric hinge, and allow for the tolerance of the printer (Flashforge Creator Pro), as well as the printing settings used to create the tiles. The additional circular openings visible in Fig.~\ref{fig:F4}D were included to enable the use of tweezers to quickly place the fabric pieces. The slots and circular pocket cut in the cad were extended to $2~mm$ above the bottom face of the tiles, leaving connecting material on the bottom for structural integrity and to hide the hinge attachment for aesthetic purposes. Similarly, a $2~mm$ cap was printed and glued in place to cover the insertion holes and slots for the fabric hinges. As a final CAD step, temporary trusses were added in the negative spaces between tiles to guide tile alignment for hinge insertion and gluing. Each truss is approximately $2~mm$ wide and $0.5~mm$ tall and extends between opposite tile faces inside of the negative spaces. The trusses are aligned with the bottom face to be printed to avoid added support material. The heart example was divided in half and the mirrored to both fit the build platform of the FFF printer used and to reduce the amount of modeling work and modification required. While this CAD preparation was done manually, it is possible to automatically generate parametric models that are then used to create an stl and tool path that could be sent to a printer.\\

Once printed, $20~mm$ strips of fabric were cut and placed inside each of the slots with the help of tweezers. The fabric was then glued in place with a few drops of Cyanoacrylate (Super Glue) placed at the intersection of the circular pocket and hinge slots with care taken to not flood the fabric exposed in the hinge with glue as this would rigidify the hinge. After the hinges were installed, the trusses were then trimmed away with an exacto blade. The caps mentioned above were then glued over the side of the tiles with the exposed hinge slots and circular pockets for aesthetic purposes. This approach yields  deployed  shapes that match the computational result very well; additionally, the resulting rigid-deployable, compact reconfigurable kirigami structures have high bending stiffness.\\

We can also create much softer kirigami sheets using a molding approach (Fig.~\ref{fig:F4}E--H). We start by producing a laser-cut acrylic molds based on a deployed configuration of the kirigami patterns obtained by our framework. A vector file of the pattern can be minimally edited in a vector editing software to add alignment pin holes, add an outer border to the mold, and prepare the file to be received by commercial laser cutters. Alternatively, processing of the patterns to a laser-cutter compatible file could also be automated with a python script. The mold could also be printed or machined, but laser cut acrylic is quick to fabricate, laser cutters are relatively prevalent in industrial and academic fabrication spaces, and the acrylic is inexpensive and compatible with the molding process. As an added benefit, the laser cut acrylic does not require mold release. The mold can be glued together but we use pins with a light press fit (less than 10 pounds of force) to both align the mold pieces and hold them together. The press fit pins are inexpensive, proved sufficiently precise locating between parts, and can be assembled by hand or with a pair of pliers. \\

Once the mold is assembled, we use tweezers to insert textile pieces into the mold between mold pieces, as shown in Fig.~\ref{fig:F4}H. These textile pieces will form the hinges. Before placing them in the mold, the textile pieces are coated in the two part liquid silicone rubber that will fill the mold to form the soft kirigami sheet in which the hinges will be embedded. The physical model in Fig.~\ref{fig:F4}F was made with Elastosil (Wacker Chemie AG, M4601AB) with cheese cloth hinges (McMaster-Carr, PN8808K11), and the physical model in process in Fig.~\ref{fig:F4}H was made with DragonSkin 20 (Smooth-On, Inc.) and generic herringbone cotton twill tape (0.25in ribbon) for the hinges. Dipping the textile hinges in rubber before installing them into the mold is to ensure saturation of the textile and full mechanical incorporation in the rubber, as well as helping to minimize bubble entrapment in the mold filling process. We install our rubber-coated fabric hinges in place and before the rubber cures, then fill the rest of the mold with a silicone rubber to produce a physical model. This approach facilitates fabrication of large arrays and allows us to minimize the size and stiffness of the hinges. For illustration, we fabricate the square-to-circle pattern obtained by nonlinear inverse design (Fig.~\ref{fig:F4}E) and see that the deployment of the physical models (Fig.~\ref{fig:F4}F) matches that of the computational results very well.

\subsubsection*{Digital preparation of 3D-printed physical models}
An overview of the process for preparing the Kirigami design for the 3D printing is shown in Supplementary Figure~20 and the post processing, hinge insertion, and assembly of the printed kirigami tiles is shown in Supplementary Figure~21. The kirigami is first saved in a vector format (such as a dxf file) to be imported into a 3D modeling software package. The result of importing a dxf file of the heart structure into a sketch within Fusion360 (Autodesk's 3D modeling software) is shown in Supplementary Figure~20A, but this approach of preparing the kirigami pattern for printing can be accomplished in various other 3D modeling software packages or automated via a Python script. The workflow that may be adapted to alternate software packages begins with a vector sketch of the kirigami pattern that is extruded perpendicularly to the sketch plane, as shown in Supplementary Figure~20B, to create a three-dimensional structure. Designs with symmetric sub-units can be divided, as shown in Supplementary Figure~20C, to later be mirrored or rotated, thus saving time and allowing larger assemblies to be made on small printer platforms.\\

For this rigid-deployable kirigami pattern, we add open slots in the extruded tiles into which textile hinges can be added after printing. These slots are indicated on an individual tile in Supplementary Figure~20D. Circular clearance pockets are added to the end of the slots in the tiles, and the pockets are large enough to allow the use of tweezers to aid in hinge insertion. Alternatively, larger portions of the tile interior can be removed, as shown in the second half of the deployable heart in Supplementary Figure~21. Removal of additional material maximizes clearance and reduces overall material usage in the printing process. For structural integrity and aesthetic purposes, however, the hinge slots and clearance pockets do not extend through the full thickness of the tiles. The thickness of the slot will depend on the hinge textile used, the printer tolerance, and the overall scale of the assembly. For this particular instantiation of the design, the tiles were $15~mm$ thick, the slots were $1.~5mm$ wide (tslot), and the height of the slots was $12~mm$ (hslot), thus leaving $3~mm$ of thickness on the bottom of the tile. The center line of the slot intersects the corner of the individual tiles, but it is not necessary for the slots to form a straight line between tiles, nor is it necessary for the slot to bisect the insertion angle of the tile corners, as is visible in Supplementary Figure~20D and Supplementary Figure~20E.\\

As a final step before printing, temporary trusses are added in the negative spaces to hold the tiles in a fixed, partially deployed configuration. As shown in Supplementary Figure~20F, the trusses hold the alignment of the tiles for hinge insertion and are later removed. While the trusses are helpful to the assembly process, they are not necessary and could be eliminated to allow for greater packing density on the printer bed. The cross-section of the trusses is approximately $0.5~mm$ $\times$ $2~mm$ and they are oriented such that they are printed against the bed, with the first few layers of the model. The trusses should be thick enough that they are not broken when the tiles are removed from the printer platform or when the hinges are installed, but they are also thin enough that they can be easily trimmed with a flush cutter or a knife.

\subsubsection*{3D-printing of physical models and hinge installation}
The prototype shown in Supplementary Figure~21 was printed on an FFF printer (Flashforge Creator pro, Zhejiang Flashforge 3D Technology Co., China), using $1.75~mm$ PLA filament (Hatchbox, California, USA), and can be easily translated to other printers and printed materials. By printing the tiles in a deployed configuration and not in either of the fully compact states, we minimize processing after printing that needs to be done to separate the tiles. An example of a single textile hinge prior to installation is shown in Supplementary Figure~21A. The length of the hinge does not need to be precise but a hinge is easier to insert with the use of one or two tweezers, in which case the hinge must be long enough to span two of the clearance pockets. After the hinges are in place and positioned as desired, superglue can be used to tack the fabric to the inside of the tiles, adding a few drops to the slot from the inside of the clearance pockets, leveraging the low viscosity of the glue to wick into the slot. Care should be taken to not get glue on the span of hinge exposed between tiles as this would rigidify the hinge textile and prevent the hinge from bending freely.

\subsubsection*{Finishing details for 3D-printed models}
For aesthetic purposes, we use a set of 3D printed caps to cover the holes and form the closed tiles of the kirigami structure. The caps are the same profile as the tiles and can be glued or press fit into place. In the example shown in Supplementary Figure~21, the caps are 2~mm thick and have 2~mm thick circular extrusions on the inner surface that line up with the clearance pockets. These extrusions are not necessary but assist with alignment. Depending on the printer and material used, as well as the press fit desired, the extrusions may vary from oversized to under-sized for the fit with the clearance pockets. In the example shown, the physical model is printed in PLA with 30 percent infill, three outer solid layers, a line-to-line fit, and a chamfer. The purpose of the chamfer is to help with insertion. The result was a light press fit capable of holding most of the caps in place. The tolerance and print quality was such that the a couple of the caps did not stay in place without light applied pressure but permanent installation of the caps was accomplished by gluing them in place and thus the tolerances were sufficient. To fix the caps in place, a bead of hot glue was piped into the clearance pockets and the cap was pressed on top. The hot glue can also reinforce the interior attachment between the hinges and tiles or may replace the super glue if there is enough contact area of the textile, glue and, and tile interior. Given the low melting temperature of PLA, care must be taken with applying hot glue. This can also be improved by printing parts in a higher temperature plastic such as Nylon, ABS or SLA resin.

\subsubsection*{Living hinges as a 3D-printed alternative}
A construction technique using living hinges could be used in place of the textile hinge installation process described above. For a living hinge, the hinge material is the same as that of the tile in a thin enough geometry that it can bend without breaking. The textile was chosen for our physical models because it is less likely to fatigue or tear while reaching a near-zero pivot radius.

\subsubsection*{Mold preparation for molded rubber physical models}
Details of the fabrication procedure of the rubber molding approach are given below, with illustrations of the mold preparation in Supplementary Figure~22 and pictures of the molding and hinge installation in Supplementary Figure~23. The mold is cut from a $6.35~mm$ ($0.25~in$) sheet of acrylic. The mold consists of two layers that are held together with $0.0625~in$ diameter stainless steel pins that are press-fit into holes that span both layers. The mold pieces could also be glued or bolted together. We find that the press fit attachment is easy to deconstruct and the ability to quickly pull rigid mold tiles from the assembly without any required tools facilitates removal from the mold after the rubber has fully cured.\\

To create a mold, we start with a vector file of the kirigami design and turn it into a vector file like the ones shown in Supplementary Figure~22A and Supplementary Figure~22B to laser cut an acrylic sheet. A shown, we use a configuration approximately halfway through the transition between fully compact states to minimize small angles and small parts that could be difficult to assemble. Laser cutting the transition state creates tiles of both the positive tile shapes and negative spaces of the four-bar linkage network that makes up the kirigami sheet. Assembling a mold of the tiles that represent negative spaces creates a positive model of the intended design when filled with a rubber or plastic. For this fabrication approach, the necessary gaps for installing hinges are created by the laser kerf and the design and dimensions are not altered from the vector pattern exported from the script described above other than uniformly scaling the entire assembly.The example molding process is shown for the physical model of the rigid-deployable square-to-circle kirigami pattern in Fig.~\ref{fig:F3}B, for which the assembly was scaled to fit within a square mold that is eight inches on a side.\\

File preparation will vary for a given laser cutter but a file setup similar to the one shown in Supplementary Figure~22A and Supplementary Figure~22B was used on both a Versalaser and an Epilog laser cutter to make suitable mold from $6~mm$ thick acrylic sheets. For this example, Adobe illustrator was used to create the vector files to run the laser cutter. There are two copies of the pattern, one for the top layer of the mold and shown in Supplementary Figure~22A and one for the bottom layer as shown in Supplementary Figure~22B. A bounding box around the pattern was added to cut the exterior border of the mold and circles were added throughout the pattern to create holes where press fit pins would hold a tile defining a negative space to the bottom of the mold. The color coding in the file allows for different power settings between cuts and control over the order in which different paths are cut. In the vector files shown, pink and red are used for through-cuts with two different speed settings, and the blue lines are etched into the acrylic as references to guide the mold assembly but are not cut through.\\

The only critical power setting to tune and test before laser cutting the mold parts is the circular holes that are used for press fit pins that hold the various part of the mold together. The press fit of the pins was designed to be tight enough so that they would not fall out if the mold is inverted but also loose enough that they could be readily pulled out without the aid of pliers. In this size range ($1/16~in$ or $1.59~mm$ diameter pins) this can be done by specifying the hole approximately \(50-200\ \mu \)m  smaller in the vector file than the physical pin diameter and adjusting the power and speed settings of the laser cuter to adjust the kerf. (To be clear, the kerf or cut width of the laser has a non-zero thickness and directing the laser around a circular path of a given diameter will result in a hole of a larger diameter than the guiding circle.) Slower cuts and higher power cuts will remove more material, creating a larger kerf, larger hole, and looser fit. We found that a test piece to calibrate the pin press-fit with several holes at increments of 25 or \( 50\ \mu \)m  was sufficient to calibrate the fit for a given laser cutter. The cut speed can also be used to fine tune the kerf and fit. The recommended order of cutting operations is for the press fit pin holes to be cut first. Otherwise, if the laser cutter bed is not flat and the tiles are cut first, the tiles may shift or drop lower once cut free from the piece of acrylic. When a tile shifts, this alters the laser focus for when the press-fit holes are then cut.\\

The kerf of a laser cutter is tapered and wider at the top. We found that this effect was not detrimental for $6.25~mm$ acrylic, but other mold fabrication methods could include milling or 3D printing if the taper proved problematic or thicker tiles would be desirable. Laser cutting was used for this work because it is a fast and inexpensive method by which to makes the molds for kirigami patterns with embedded textile hinges. Additional layers could be added to a laser cut mold for thicker tiles or added complexity. Additionally, while we demonstrate the creation of soft kirigami sheets, a similar process could be used to mold a rigid plastic with textile hinges. The mold would need to be modified to facilitate the release of rigid molded tiles. This can be accomplished with added draft angles, which may be accomplished with the laser kerf, or the mold could be made of a soft material.\\

As mentioned above, the blue lines visible in Supplementary Figure~22A and Supplementary Figure~22B are etched and not cut through the acrylic sheet. These lines may be omitted but provide visual guidelines for assembling the mold. A partially assembled mold is shown in Supplementary Figure~22C, where it is difficult to see the etched lines on the white acrylic. To make the lines more easily visible, ink can be wiped into the etched lines, as shown in Supplementary Figure~22D. The ink used in this case was a silicone dye (Silc Pig, Smooth-on, Inc.) and the excess was wiped away with a paper towel. The pin holes provide mechanical alignment and the etched numbers and lines help to visually guide manual assembly of the mold, as shown in Supplementary Figure~22E. The modular press-fit mold is useful for small scale manufacturing but, ultimately, this could be converted to a two-part injection mold for higher through-put fabrication.

\subsubsection*{Hinge installation and molding soft kirigami sheets}
Once fully assembled, the acrylic mold is ready for rubber casting and does not require mold release because silicone rubber will not bond to the acrylic sheet or stainless steel pins. The molding process is shown in Supplementary Figure~23, starting with preparation of the textile in Supplementary Figure~23A that is used to create the hinges of the kirigami pattern. The pattern can be molded out of rubber alone (without textile hinges) but would then rip very easily. The textile reinforces the hinge while still allowing a full range of motion and transformation of the mechanism. The fabric used in the rigid-deployable square-to-circle kirigami model in Fig.~\ref{fig:F4} was a cotton cheese cloth, while the fabric used for the model in Supplementary Figure~23 is a herringbone cotton twill tape. The cheese cloth is cut into approximately 12~mm squares and folded in half (creating rectangles) before insertion into the mold. The twill tape is approximately 6mm wide and thus only needs to be cut to length to fit into the mold of 6.35~mm acrylic. The cotton fibers in either choice of textile have a high surface area and create a strong mechanical bond with the silicone. Synthetic fibers will work but were found delaminate more readily from the cast silicone. As shown in Supplementary Figure~23B, the twill tape is prepared by cutting it into pieces approximately 6-12~mm in length, long enough to span into neighboring cast tiles.\\ 

A two part silicone rubber is mixed and then put under vacuum to remove bubbles. While the rubber is degassing, a small portion was set aside, as visible in Supplementary Figure~23B, to start coating the cut pieces of twill tape. By coating textile hinges (twill tape or cheese cloth) before inserting them, we found we could improve the saturation of the rubber in the textile, reduce bubbles, and improve mechanical integration between the textile and rubber. The rubber used in the model shown in Fig.~\ref{fig:F4} was a two part silicone rubber, Elastosil m4601 A/B (Wacker, Munich, Germany). The rubber used in Supplementary Figure~23 is Dragonskin20 (SmoothOn, Inc. with blue dye added (Silc Pig, SmoothOn, Inc.) to help monitor the mixing and improve visibility in pictures for documentation. Silicone is an appealing material to use because of its durability, elastic strain behaviour, and ease of use, but a large variety of rubbers would work to create these structures. A soft silicone rubber is also relatively easy to pull out of the rigid mold and does not require mold release. A rigid plastic or urethane could also be cast molded with relatively minor mold modifications.\\

Once coated with rubber, the textile pieces are inserted into the mold to form the hinges (Supplementary Figure~23C) and covered with a liquid rubber to form the kirigami structure (Supplementary Figure~23D--E). In the example shown (Supplementary Figure~23C), the textile is pinched between the corners of the acrylic mold tiles that define the open spaces of the kirigami structure to create a  cotton and silicone composite hinge. The laser kerf is sufficient to create a gap between the mold tiles for the textile. Two pairs of tweezers, one on either side of the hinge, are very helpful for the process of inserting the textile, as shown in Supplementary Figure~23C. The hinges can be inserted before or after the rubber is poured into the mold, as shown in Supplementary Figure~23C--E. Once the rubber is fully cured, the soft krigiami sheet can be peeled out of the mold and trimmed of excess rubber, as shown in Supplementary Figure~23F. If the excess rubber is not trimmed, it can over constrain the kirigami sheet, preventing the hinges from pivoting through the intended motion. The excess rubber can also be prevented with careful pouring, by wiping the excess before curing, or by creating a fully enclosed mold with a precise or compliant interface. The reconfiguration of the hinges can be seen in the translucent soft kirigami sheet as it is transformed from a square in Supplementary Figure~23G, to a mid-deployment state in Supplementary Figure~23H, and to a circle in Supplementary Figure~23I.


\bibliographystyle{ieeetr}

\begin{figure}[t]
\centering
\includegraphics[width=0.8\textwidth]{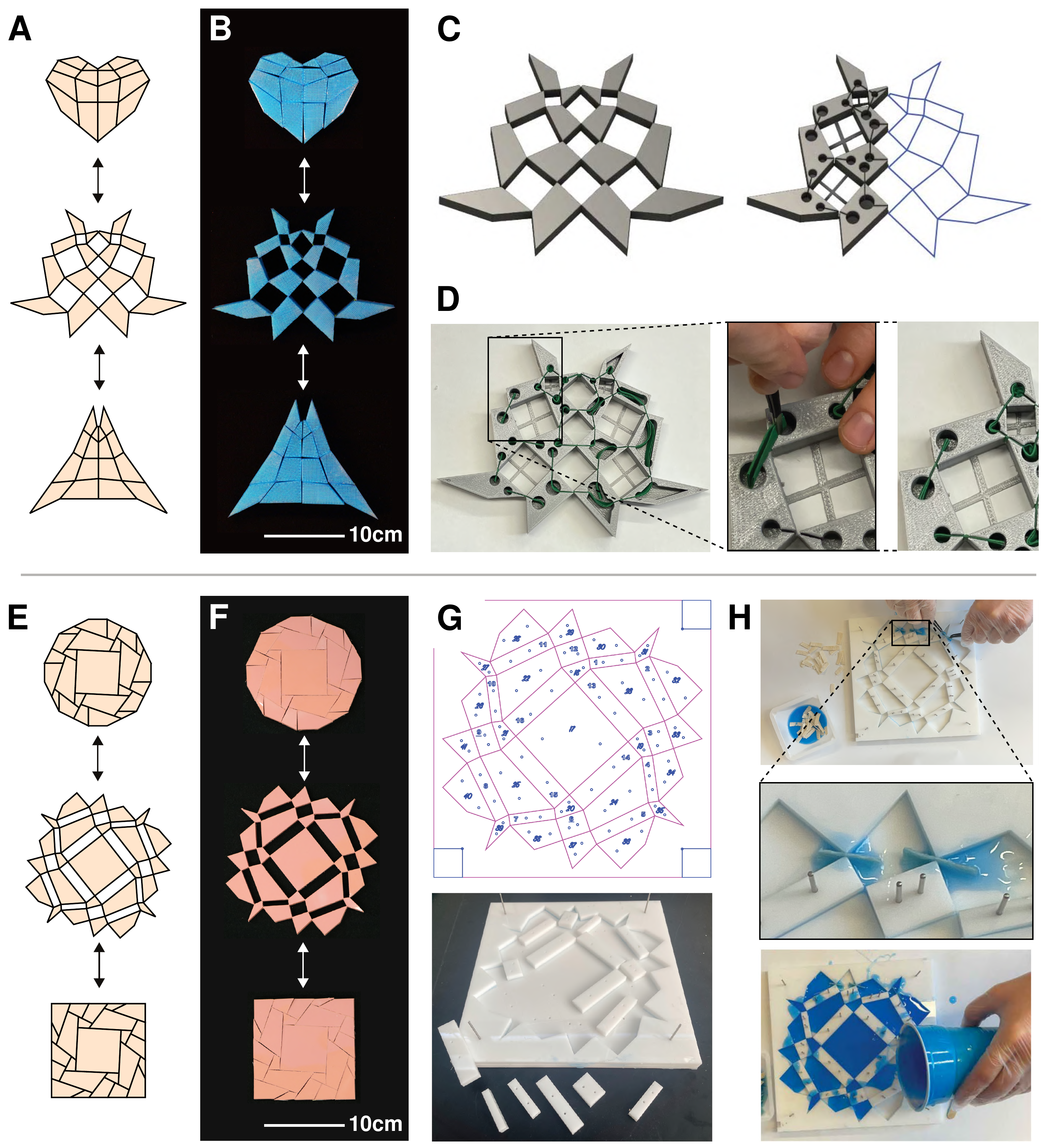}
\caption{\textbf{Physical model fabrication for rigid-deployable kirigami fabricated via a 3D printing based method and a cast silicone composite method.} \textbf{(A)} Snapshots of the deployment of the heart pattern in Fig.~\ref{fig:F2}C and \textbf{(B)} a corresponding 3D-printed model. \textbf{(C)} A vector file of the pattern mid-deployment imported into 3D modeling software and extruded to a chosen thickness. Slots are digitally cut out for installing textile hinges, and temporary trusses added to fix deployment angles for hinge installation. Larger holes make it easier to insert hinges and patterns can be processed in symmetric sub units for efficiency and later mirrored or rotated. \textbf{(D)} Once the rigid tiles are 3D printed, fabric hinges are inserted in the slots using tweezers as shown in the zoom-in image. The hinge insertion holes can later be covered for aesthetic purposes. \textbf{(E)} Snapshots of the deployment of the circle-square pattern in Fig.~\ref{fig:F2}C and \textbf{(F)} a corresponding pink elastomeric model. \textbf{(G)} A laser cut mold based on a deployed configuration is created with the vector file of the pattern shown in the top panel. \textbf{(H)} Fabric pieces are inserted in the gaps between mold pieces using tweezers (top) and the mold is filled with blue silicone rubber to form the final soft kirigami sheet (bottom).}
\label{fig:F4}
\end{figure}

\end{document}